\documentclass[twocolumn]{aastex6}
\usepackage{amsmath}
\usepackage{enumerate}
\usepackage{rotating}

\def\numax{$\nu_{\rm max}$}

\def\dnu{$\Delta\nu$}

\begin{document}

\title{Determining the Best Method of Calculating the Large Frequency Separation For Stellar Models}

\author{Lucas S. Viani\altaffilmark{1}, Sarbani Basu\altaffilmark{1}, Enrico Corsaro\altaffilmark{2}, Warrick H. Ball\altaffilmark{3,4}, and William J. Chaplin\altaffilmark{3,4}}
 
\altaffiltext{1}{Department of Astronomy, Yale University, New Haven, CT, 06520, USA}
\altaffiltext{2}{INAF$–-$Osservatorio Astrofisico di Catania, via S. Sofia 78, 95123 Catania, Italy}
\altaffiltext{3}{School of Physics and Astronomy, University of Birmingham, Edgbaston, Birmingham, B15 2TT, UK}
\altaffiltext{4}{Stellar Astrophysics Centre (SAC), Department of Physics and Astronomy, Aarhus University, Ny Munkegade 120, DK-8000 Aarhus C, Denmark}

\email{lucas.viani@yale.edu}

\begin{abstract}
Asteroseismology of solar-like oscillators often relies on the comparisons between stellar models and stellar observations in order to determine the properties of stars. The values of the global seismic parameters, \numax\, (the frequency where the smoothed amplitude of the oscillations peak) and \dnu\ (the large frequency separation), are frequently used in grid-based modeling searches. However, the methods by which \dnu\ is calculated from observed data and how \dnu\ is calculated from stellar models are not the same. Typically for observed stars, especially for those with low signal-to-noise data, \dnu\ is calculated by taking the power spectrum of a power spectrum, or with autocorrelation techniques. However, for stellar models, the actual individual mode frequencies are calculated and the average spacing between them directly determined. In this work we try to determine the best way to combine model frequencies in order to obtain $\Delta \nu$ that can be compared with observations. For this we use stars with high signal-to-noise observations from \textit{Kepler} as well as simulated TESS data of \cite{Ball2018}. We find that when determining \dnu\ from individual mode frequencies the best method is to use the $\ell=0$ modes with either no weighting or with a Gaussian weighting around \numax.
\end{abstract}


\section{Introduction}

In the field of asteroseismology, stellar models play a key role in determining the properties of observed stars. Just by knowing the basic seismic parameters, \numax\ (the frequency where the smoothed amplitude of the oscillations peak) and \dnu\ (the large frequency separation), as well as $T_{\mathrm{eff}}$ for a star, models can be used to place constraints on stellar age, radius, and mass. Since there is such frequent reliance on matching the seismic parameters determined from observations to the seismic parameters extracted from stellar models, we need to be sure that the methods by which we calculate \dnu\ and \numax\ from models produce accurate representations of the observed global values. Determining the correct way to extract the value of \dnu\ and \numax\ from a stellar model is therefore of great importance. 

The large frequency separation, \dnu, is the average frequency spacing between modes of adjacent radial order ($n$), of a given degree ($\ell$). The radial order $n$ is the number of nodes in the radial direction and $\ell$ is the number of node lines on the star's surface. This quantity $\Delta \nu$ arises from the asymptotic relation \citep{Tassoul1980,Gough1986}, which is applicable for modes of low $\ell$ and high $n$. The relation is not exact and the spacings between the modes have some variability. Therefore, the value of $\Delta \nu$ will depend on the method by which this average spacing is calculated.

The value of \dnu\ can be approximately related to the density of the star, as $\Delta \nu \propto \sqrt{\bar{\rho}}$ \citep[see, e.g.][]{Tassoul1980,Ulrich1986,CD1988,CD1993}. This leads to the \dnu\ scaling relation,
\begin{equation}
\frac{\Delta \nu}{\Delta \nu_{\odot}} \simeq \sqrt{\frac{M/M_{\odot}}{(R/R_{\odot})^{3}}}.
\label{eq:dnu_scaling}
\end{equation}
For an observed star, the value of \dnu\ can be determined from the excited p-modes in the star's power spectrum. This observed \dnu\ is not usually calculated directly from individual mode spacings, as this would require high signal-to-noise, but instead through other methods. For example, \dnu\ is often determined by taking a power spectrum of a power spectrum \citep{Mathur2010,Hekker2010} or autocorrelation techniques \citep{Roxburgh2006,Roxburgh2009,Mosser2009,Huber2009,Rene2013,Verner2011}.

It is important to note that this is not how the value of \dnu\ is determined for stellar models. Since for a stellar model the radius and mass are known quantities, a simple approach to determine \dnu\ would be to make use of the \dnu\ scaling relation in Eq.~\ref{eq:dnu_scaling}. However, studies have shown that the \dnu\ scaling relation has deviations, is a function of $T_{\mathrm{eff}}$ and $\mathrm{[Fe/H]}$, and only holds to a few percent \citep{White2011, Mosser2013, Miglio2013, Guggenberger2016, Sharma2016, Yildiz2016, Rodrigues2017, Ong2019}. Therefore, \dnu\ for stellar models is not usually determined using the scaling relation, but by calculating the model's individual mode frequencies. The value of \dnu\ can then be calculated by finding the average spacing between these frequencies (of a certain $\ell$). This average is usually determined as the slope of a linear fit to the $\nu$--$n$ relationship for modes of a given $\ell$. Since the spacing of modes is not exactly the same throughout the excited mode envelope, then the manner in which the averaging is performed is important. This leaves some ambiguity as to what the best method of calculating \dnu\ from these individual mode frequencies is. For example, which $\ell$ modes to include in the averaging or whether to weight the modes around \numax\ more heavily, are decisions which can produce important differences in the value of \dnu. \cite{Rodrigues2017} shows that there is a difference between Gaussian weighted and error weighted values of \dnu\ up to about 1\%. \cite{Roxburgh2014} also discusses that depending on the method of calculating \dnu, the results can differ by about 1\%. If relying on the scaling relation to determine mass, a 1\% deviation in \dnu\ can have a meaningful impact.

The frequency where the smoothed amplitude of the oscillations peak, \numax, can be shown to be proportional to the acoustic cutoff frequency, $\nu_\mathrm{ac}$ \citep{Belkacem2011}, and goes as $\nu_{\mathrm{max}} \propto \nu_\mathrm{ac} \propto g T_{\mathrm{eff}}^{-1/2}$ \citep{Brown1991,Kjeldsen1995,Bedding2003,Belkacem2011}. This leads to the \numax\ scaling relation, given by
\begin{equation}
\frac{\nu_{\mathrm{max}}}{\nu_{\mathrm{max,\odot}}} \simeq \left(\frac{M}{M_{\odot}}\right)  \left( \frac{R}{R_{\odot}}\right)^{-2}  \left(\frac{T_{\mathrm{eff}}}{T_{\mathrm{eff,\odot}}}\right)^{-1/2}.
\label{eq:numax_scaling}
\end{equation}
While one could determine the approximate value of \numax\ for a stellar model using Eq.~\ref{eq:numax_scaling}, the \numax\ scaling relation must be used with caution, as many studies have shown that deviations do exist \citep{Bedding2003, Stello2009, Bruntt2010, Miglio2012, Bedding2014, Coelho2015, Silva2015, Yildiz2016, Viani2017}. For stellar models, a more accurate way to calculate \numax\ is to avoid the \numax\ scaling relation and instead use the acoustic cutoff frequency, as described in \cite{Viani2017}.

In this paper we investigate various methods of calculating the large frequency separation from individual mode frequencies (referred to as $\Delta \nu_\mathrm{freq}$ from now on) to determine which method gives values of \dnu\ that best represent the global observational \dnu\ values. This will be done using observations of high signal-to-noise stars, with individual mode frequencies already determined from previous studies in the literature. For these stars, the individual mode frequencies will serve as a proxy for the frequency values that one would have if they were modeling the star. For each star in our sample, the value of $\Delta \nu_\mathrm{freq}$ will be calculated using a variety of methods and compared to the global value of $\Delta \nu$, calculated using standard observational methods. This will allow us to determine the optimal way to calculate \dnu\ from individual mode frequencies and develop a better understanding of how $\Delta \nu_\mathrm{freq}$ of stellar models should be determined. We then verify these results using the \cite{Ball2018} simulations of lightcurves for NASA's \textit{Transiting Exoplanet Survey Satellite}, TESS \citep{Ricker2015}. This issue is especially important as a multitude of new observations from TESS become available.  

The paper is organized as follows, Sec.~\ref{Sec:Data} gives an overview on the stars used in the study, explains the methods used to determine the seismic parameters from the observed data, and discusses the various methods used to calculate \dnu\ from the individual mode frequencies. Sec.~\ref{Sec:Results} presents the results and compares the different values of $\Delta \nu$. Sec.~\ref{Sec:Discussion}  discusses the findings and Sec.~\ref{Sec:Conclusion} provides concluding remarks.

\section{Data and Analysis}
\label{Sec:Data}
\subsection{Sample of Stars in the Study}

The stars used in this study consist of the 66 main sequence stars from the \textit{Kepler} Asteroseismic LEGACY Sample from \cite{Lund2017}, 34 solar-type planet-hosting stars from \cite{Davies2016}, the 23 main sequence and subgiant stars from \cite{App2012} that were not already included from the LEGACY Sample, and 17 red giant stars from NGC 6791 that were in \cite{Corsaro2017} and \cite{McKeever2019}. It should be noted that while \cite{Davies2016} examined 35 stars, we excluded KIC 8684730 as it did not have a readily available multi-quarter power spectrum. For each of the 140 stars in our sample, the \textit{Kepler} power spectrum was obtained from the KASOC website\footnote{kasoc.phys.au.dk}. For the main sequence and subgiant stars the short cadence KASOC weighted version of the power spectra were used while for the RGB stars the ``Working Group 8'' long cadence data were used. The seismic parameters were extracted from the power spectrum using several different methods as described in Sec~\ref{Sec:Determining_Obs_Dnu}.

\subsection{Determining Seismic Parameters from Power Spectra}
\label{Sec:Determining_Obs_Dnu}
\subsubsection{2D Autocorrelation Method}
\label{Sec:ACF}
\begin{figure}
\begin{center}
\includegraphics[width=7cm]{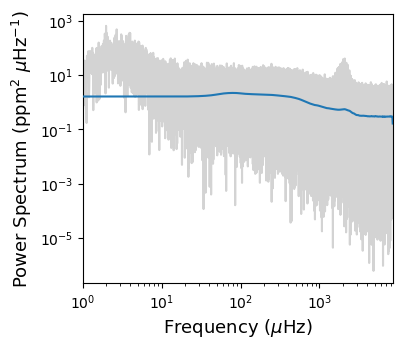}
\end{center}
\caption{The power spectrum for KIC 6116048. The blue line shows the smoothed background estimate.}
\label{fig:PS_smoothed}
\end{figure}

One method to determine the value of \numax\ and $\Delta \nu$ from a power spectrum is the 2D autocorrelation function (ACF) method, as in \cite{Huber2009}, \cite{Verner2011}, and \cite{Rene2013}. The premise of this technique is to perform a series of autocorrelations on segments of the power spectrum to determine the frequency range of the envelope of excited modes. First the power spectrum is smoothed, to estimate the background, using a median filter with a window size of 100 $\mu$Hz for the main sequence and subgiant stars and a window of 10 $\mu$Hz for the giants. An example of this smoothing for star KIC 6116048 can be seen in Fig.~\ref{fig:PS_smoothed}. Then the power relative to the smoothed background spectrum (PBS) is determined using Eq.~\ref{eq:PSB} \citep{Verner2011, Rene2013}
\begin{equation}
\label{eq:PSB}
PBS(\nu) = \frac{P(\nu) - Bg(\nu)}{Bg(\nu)},
\end{equation}
where $P(\nu)$ is the power at a given frequency and $Bg(\nu)$ is the smoothed background.

\begin{figure}[h]
\begin{center}
\includegraphics[width=7cm]{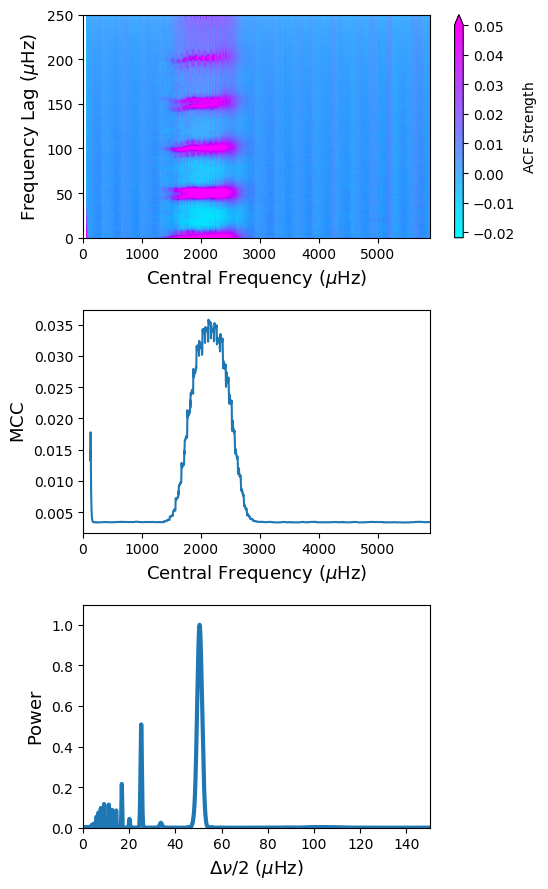}
\end{center}
\caption{\textbf{Top:} The 2D autocorrelation results for KIC 6116048. The abscissa shows the central window frequency, the ordinate shows the autocorrelation lag, and the colors indicate the autocorrelation power. \textbf{Middle:} The mean collapsed correlation (MCC). The abscissa shows the central window frequency and the ordinate shows the average absolute value of the autocorrelation power for each window (see Eq.~\ref{eq:MCC}). \textbf{Bottom:} The power as a function of $\Delta \nu /2$ for the PS$\otimes$PS of the p-mode envelope for the example star KIC 6116048.}
\label{fig:ACF_figure}
\end{figure}

A series of autocorrelations are then performed on different segments of the PBS. Starting with the lowest frequency in the power spectrum, an autocorrelation is calculated for a 250 $\mu$Hz wide window (25 $\mu$Hz window for the giant stars). The window size for the main sequence stars was chosen to match that of \cite{Rene2013}. The central frequency of the window is then shifted by 1 $\mu$Hz (as in \cite{Verner2011} and \cite{Rene2013}), towards higher frequencies, and another autocorrelation is performed. This is continued until the window reaches the end of the power spectrum. The results of this process can be seen in the top pannel of Fig.~\ref{fig:ACF_figure}, where the autocorrelation power for each frequency lag can be plotted as a function of the central window frequency.

As can be seen in the top pannel of Fig.~\ref{fig:ACF_figure}, when in the frequency range of the excited p-modes, the autocorrelation power spikes with a regular spacing which corresponds to $\Delta \nu /2$. This clear pattern in the autocorrelation power is not present outside of the frequency range of the p-mode envelope. Thus, by examining where the autocorrelation shows this spacing, we can determine the frequency range of the envelope of excited modes as well as the value of $\nu_{\mathrm{max}}$. 

To make this more clear, we can collapse the top panel in Fig.~\ref{fig:ACF_figure} to examine just the total average autocorrelation power at each central frequency. This quantity, called the mean collapsed correlation (MCC: \cite{Rene2013}), is calculated for each central frequency as,
\begin{equation}
\mathrm{MCC} = \frac{(\sum_{i=1}^{n_\mathrm{lags}} |ACF_{i}|)-1}{n_\mathrm{lags}}
\label{eq:MCC}
\end{equation}
where $n_\mathrm{lags}$ is the number of lags in the autocorrelation function. In the numerator 1 is subtracted because at lag 0 the autocorrelation is 1 since the spectrum has not been shifted. The presence of the absolute value in the equation is because a negative correlation power also holds valuable information. We plot the MCC as a function of central window frequency in the middle panel of Fig.~\ref{fig:ACF_figure}, again for KIC 6116048.
 
From the collapsed 2D autocorrelation, the frequency range in which the excited p-modes reside can clearly be seen. A Guassian is then fit to the MCC peak, with the Gaussian's center being \numax, as done for example by \cite{Huber2009}, \cite{Verner2011}, and \cite{Rene2013}. The envelope of excited p-modes is then defined, following \cite{Rene2013}, to be the frequency range around \numax\ where the MCC value is at least 10\% of the Gaussian peak height.

Once the frequency range of the excited p-mode envelope is determined, the value of \dnu\ can be calculated by taking a power spectrum of the power spectrum (PS$\otimes$PS) for this frequency range \citep[see, e.g.,][]{Mathur2010,Hekker2010}. A Lomb-Scargle periodogram \citep{Lomb1976,Scargle1982} is computed on the PBS spectrum for the frequency range of the excited envelope. For the example star, KIC 6116048, the PS$\otimes$PS can be seen in the bottom panel of Fig.~\ref{fig:ACF_figure}. A Gaussian is then fit to the periodogram, with the Gaussian's center corresponding to $\Delta \nu /2$. The $\ell=1$ peak falls between the $\ell=0$ peaks and so the power maximizes at $\Delta \nu / 2$ instead of at \dnu. To help fit the Gaussian, and determine the correct peak in the PS$\otimes$PS, we estimate $\Delta \nu_{\mathrm{expected}}$ using our calculated value of \numax. Many studies have shown a relationship where $\Delta \nu \propto \nu_\mathrm{max} ^ \beta$ where $\beta$ is between about 0.7 and 0.8  \citep[see, e.g.,][]{Hekker2009,Stello2009,Huber2011,Yu2018}, which allows us to determine which of the peaks in the PS$\otimes$PS corresponds to $\Delta \nu /2$. The uncertainty in the location of the peak is determined using the standard deviation of grouped data as in \cite{Hekker2010},
\begin{equation}
s=\sqrt{\frac{\sum fx^2 - \frac{(\sum f x)^2}{\sum f}}{\sum f -1}}
\label{eq:grouped_data_error}
\end{equation}
where f is the bin height, x is the frequency, and the summation includes the bins around the peak which have a height $\ge$1\% of the peak's height. In the remainder of the paper we will refer to the seismic parameters determined using the ACF method as $\nu_{\mathrm{max,ACF}}$ and $\Delta \nu_{\mathrm{ACF}}$.

\subsubsection{Determining Seismic Parameters using the Coefficient of Variation Method}
\label{Sec:CV_method}
While the 2D autocorrelation method has been shown to provide reliable measurements of seismic parameters it can be computationally time consuming. To obtain another set of \numax\ and \dnu\ measurements for our data we also determined the seismic parameters using a more efficient technique. The recent work of \cite{Bell2018} has shown that \numax\ can be quickly determined using what is called the coefficient of variation, or CV. The coefficient of variation is the ratio of the standard deviation to the mean of the power spectrum. The basic premise is that in a power spectrum of pure noise, this ratio should be about 1. Thus, examining where in the power spectrum this ratio is greater than 1 can be used to determine the location of solar-like oscillations.

\begin{figure}
\epsscale{1.0}
\plotone{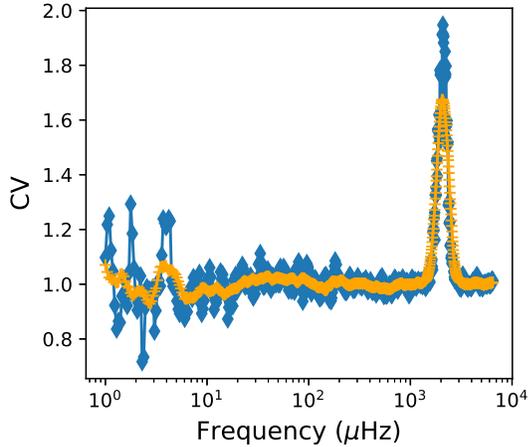}
\caption{The coefficient of variation for KIC 6116048. The blue diamonds show the CV value for each window and the yellow points show the smoothed trend.}
\label{fig:CV_plot}
\end{figure}

Our implementation of the CV method to determine \numax\ and the frequency range of the excited p-mode envelope is as follows. First, the power spectrum is broken up into a series of segments and the CV value is calculated for each segment. Starting with a window centered at 1 $\mu$Hz, the window size is set to be the same as the estimated value of \dnu\ if the central frequency were assumed to be the value of \numax. This is done to ensure that the window size is large enough so that if there were oscillations present, some would fall within the window. The $\Delta \nu_{\mathrm{estimate}}$ value is calculated assuming that $\Delta \nu_{\mathrm{estimate}}=0.267  \nu_{\mathrm{max}}^{0.764}$ as in \cite{Yu2018}. With the window size for this central frequency defined, the CV ratio is calculated for this window. The central frequency then shifts to higher frequencies by 1/6 of the previous window size. A new window size is calculated based on the new central frequency and the CV value is found again. The process is repeated until the end of the power spectrum is reached. The CV value for each window can be seen as the blue diamond points in Fig.~\ref{fig:CV_plot}. 

Next, the CV values from the different overlapping windows are smoothed. For each central frequency (each blue point in Fig.~\ref{fig:CV_plot}) the width of the smoothing window is given by $0.66 \nu_{\mathrm{central}}^{0.88}$ (referred to as $W_{\mathrm{Mosser}}$ in the remainder of the paper), based on the FWHM of the excited mode envelope from \cite{Mosser2012}. The CV values within this window are then averaged together. The resulting smoothed CV trend can be seen as the yellow points in Fig.~\ref{fig:CV_plot}. The location of the highest smoothed value (the highest yellow point in Fig.~\ref{fig:CV_plot}) is used as the initial estimate of \numax. From this initial $\nu_{\mathrm{max,estimate}}$, a weighted mean is performed, using the points that are within a window of $0.66 \nu_{\mathrm{max,estimate}}^{0.88}$, to determine the true value of \numax. The weighted mean of the peak is calculated by,
\begin{equation}
\mathrm{Peak \ \ Centroid} = \frac{\sum_{i=1}^{j} \nu_{i} \times CV(\nu_{i})}{\sum_{i=1}^{j} CV(\nu_{i})}
\end{equation}
where $\nu$ is the frequency, $CV(\nu)$ is the smoothed CV value at that frequency, and $j$ is the number of frequency bins that are within the window defined by $0.66 \nu_{\mathrm{max,estimate}}^{0.88}$. The uncertainty in \numax\ was again calculated using the standard deviation of grouped data, Eq.~\ref{eq:grouped_data_error}.

The value of \dnu\ was then calculated in a manner similar to the method used in the 2D autocorrelation approach. A Gaussian was fit to the smoothed CV \numax\ peak and the envelope of excited p-modes was determined to be the frequency range where the value was at least 10\% of the peak height. A Lomb-Scargle periodogram was then computed on the power spectrum for this frequency range, following the traditional PS$\otimes$PS method of determining \dnu\ \citep[see, e.g.,][]{Mathur2010,Hekker2010}. As previously described in Sec.~\ref{Sec:ACF}, a Gaussian was then fit to the peak in the periodogram, were the Gaussian's center corresponds to $\Delta \nu /2$. As before, the uncertainty in the location of the peak is determined using the standard deviation of grouped data as in \cite{Hekker2010}. The seismic parameters determined using the CV method will be referred to as $\nu_{\mathrm{max,CV}}$ and $\Delta \nu_{\mathrm{CV}}$ for the remainder of the paper.

It should be noted that our implementation of the CV method is not identical to that of \cite{Bell2018}. This is necessary because the \cite{Bell2018} CV method was designed for red giant stars, using long-cadence light curves, while our sample also contains main sequence and subgiant stars that have short-cadence data. For example, \cite{Bell2018} use 2000 overlapping bins spaced evenly in log-frequency (for their ``oversampled'' spectrum) while we implement the moving window overlapping by an amount based on the previous window size. This allows our windows to behave in the same manner regardless of whether we are using long or short-cadence data. Since our implementation is different, the parameter choices which we used were tested (see Appendix~\ref{Sec:Appendix}) to ensure that we were determining the location of \numax\ correctly. For example, we examine the impact of changing the window size and smoothing size. As can be seen in Appendix~\ref{Sec:Appendix}, the CV method described in this section provided the best \numax\ values.

\subsection{Comparing $\Delta \nu$ Results}
\begin{figure*}
\epsscale{1.0}
\plotone{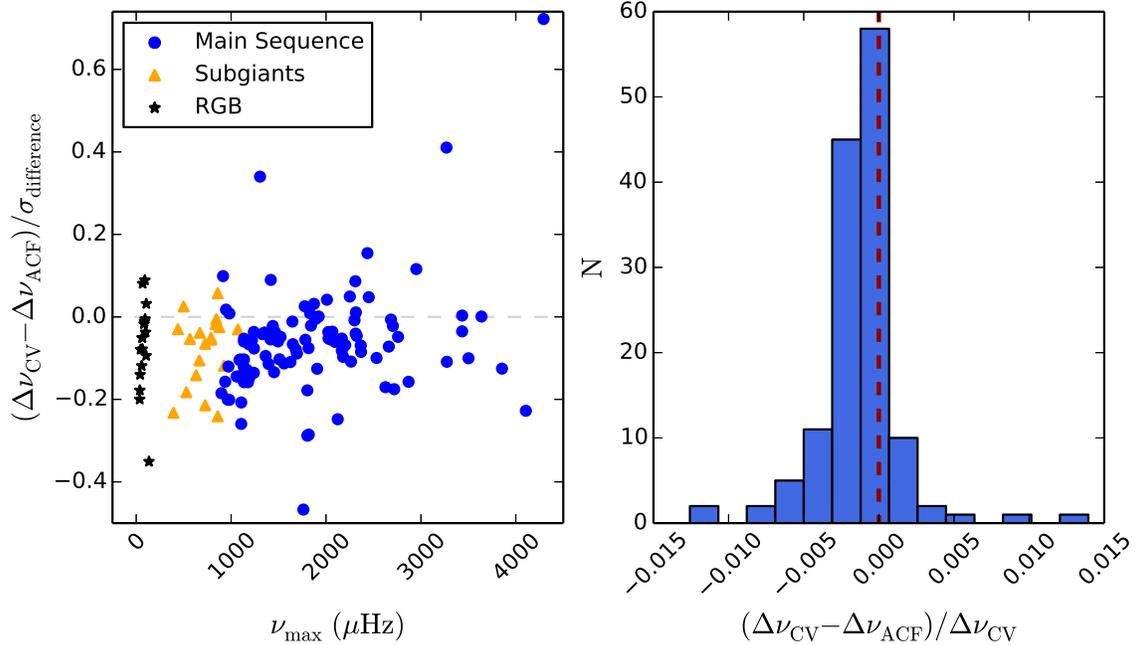}
\caption{The difference between $\Delta \nu_{\mathrm{CV}}$ and $\Delta \nu_{\mathrm{ACF}}$ normalized by the uncertainty, $\sigma_\mathrm{difference}$ (left). $\sigma_\mathrm{difference}$ is the propagated uncertainty for the value of $\Delta \nu_\mathrm{CV}-\Delta \nu_\mathrm{ACF}$. The black points show the RGB stars, the orange triangles are the subgiants, and the blue circles are the main sequence stars. The right panel shows a histogram of the fractional difference between $\Delta \nu_{\mathrm{CV}}$ and $\Delta \nu_{\mathrm{ACF}}$ for all the stars in the sample. The red dashed line marks 0 difference.}
\label{fig:ACF_Bell_dnu_compare_plot}
\end{figure*}

\begin{figure}
\epsscale{1.0}
\plotone{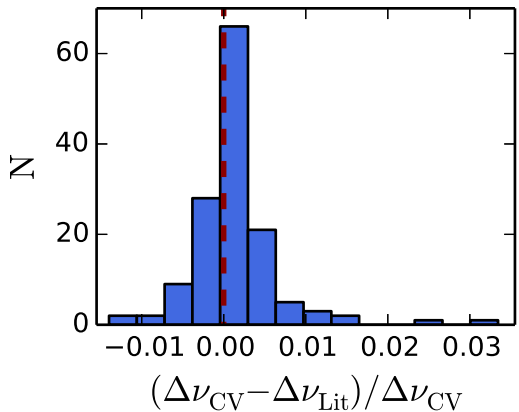}
\caption{The fractional difference between $\Delta \nu$ calculated using the CV method and $\Delta \nu$ from the literature for our sample of stars. The red dashed line marks 0 difference.}
\label{fig:dnu_compare_to_paper}
\end{figure}

Since the ACF and CV methods define the frequency range of the excited p-mode envelope slightly differently, the resulting \dnu\ value from the PS$\otimes$PS will be affected. Figure \ref{fig:ACF_Bell_dnu_compare_plot} shows the difference between $\Delta \nu_{\mathrm{CV}}$ and $\Delta \nu_{\mathrm{ACF}}$. As can be seen in Fig.~\ref{fig:ACF_Bell_dnu_compare_plot}, the value of $\Delta \nu_{\mathrm{CV}}$ tends to be slightly lower than the value of $\Delta \nu_{\mathrm{ACF}}$, however, the values of \dnu\ from the two methods are in excellent agreement, with the difference being less than 0.5\% for the vast majority of stars, and less than 1.5\% in all cases. The spread in the difference between the values of $\Delta \nu_{\mathrm{CV}}$ and $\Delta \nu_{\mathrm{ACF}}$ is less than the spread when comparing our $\Delta \nu$ values to those in the literature (as seen in Fig.~\ref{fig:dnu_compare_to_paper}). It should also be noted that the CV method offers a significant speed advantage over the ACF method.

Additionally, we can compare the value of $\Delta \nu$ determined using the CV method to the value of $\Delta \nu$ from the literature for our set of stars. For the comparison, values of $\Delta \nu$ for our sample of stars were obtained from \cite{Lund2017}, \cite{Davies2016}, \cite{App2012}, \cite{Huber2013}, \cite{Bellamy2015}, and \cite{BellamyThesis}. A histogram of the fractional difference between our calculated value of $\Delta \nu_\mathrm{CV}$ and the corresponding literature value of $\Delta \nu$ can be seen in Fig.~\ref{fig:dnu_compare_to_paper}. As can be seen in Fig.~\ref{fig:dnu_compare_to_paper}, our calculated values of $\Delta \nu$ agree very well with the literature values of $\Delta \nu$. Over 94\% of the stars have calculated values of $\Delta \nu$ within 1\% of their corresponding $\Delta \nu$ value in the literature and the distribution is centered around zero. Even in the most extreme case, the difference in the value of $\Delta \nu$ is around 3\%.

\subsection{Testing the Seismic Parameter Extraction on Solar Data}
\label{Sec:Solar_test}

Both the 2D autocorrelation method and the coefficient of variation method to extract seismic parameters were then tested using solar data. This exercise served as a check to ensure our methods of seismic parameter extraction were providing reasonable values for $\Delta \nu$ and $\nu_\mathrm{max}$ and as a way to compare the two methods. The Solar data was obtained from the 1 minute cadence photometric observations from the VIRGO instrument \citep{VIRGO1995,VIRGO1997} on the ESA/NASA spacecraft \textit{SOHO}. A segment of VIRGO data was used that was the same length as our \textit{Kepler} data. Using the CV method, the value of \numax\ and \dnu\ for the Sun were 3091.4$\pm$11.9 $\mu$Hz and 135.0$\pm$1.4 $\mu$Hz while for the 2D autocorrelation method the value of \numax\ and \dnu\ were 3417.6$\pm$81.6 and 135.5$\pm$1.5 $\mu$Hz. The typical accepted values of $\nu_{\mathrm{max,\odot}}$ and $\Delta \nu_{\odot}$ are 3090 and 135.1 $\mu$Hz \citep[e.g., see][]{Huber2011}. The value of $\Delta \nu_{\odot}$ for both the 2D ACF method and the CV method are in good agreement with the accepted value of $\Delta \nu_{\odot}$, with the CV method's value being slightly closer to the accepted $\Delta \nu_{\odot}$. Looking at the $\nu_{\mathrm{max,\odot}}$ value, while the CV $\nu_{\mathrm{max,\odot}}$ value is in good agreement with the accepted value, the 2D ACF $\nu_{\mathrm{max,\odot}}$ value is too large. While the focus of the paper is on measurements of \dnu, obtaining the correct value of \numax\ is important as the value of \numax\ can in some cases affect the calculation of $\Delta \nu_{\mathrm{freq}}$ (see Sec.~\ref{Sec:dnu_freq_calc}). Therefore, due to our inability to reproduce $\nu_{\mathrm{max,\odot}}$ using our ACF prescription, along with the fact that it is time consuming, we use the seismic parameters determined using the CV method for the remainder of this work.

\subsection{Determining $\Delta \nu$ from Individual Mode Frequencies}
\label{Sec:dnu_freq_calc}
Since the goal of this work is to compare the value of $\Delta \nu_{\mathrm{CV}}$ to $\Delta \nu_{\mathrm{freq}}$, we also must calculate the large separation using the individual mode frequencies for these stars. Observed individual mode frequencies for the \cite{Lund2017}, \cite{Davies2016}, and \cite{App2012} stars were obtained from the corresponding publications. For the NGC 6791 red giant stars the individual mode frequencies were determined by peak-bagging using the \textsc{Diamonds} code\footnote{Software and \textsc{Diamonds} code description are available at https://github.com/EnricoCorsaro/DIAMONDS} \citep{Corsaro2014} and the methodology for red giants \citep{Corsaro2015}, for a sample of cluster red giants from \cite{Corsaro2017}. The \textsc{Diamonds} code determines parameters using a nested sampling Monte Carlo algorithm. 

From the individual mode frequencies $\Delta \nu_\mathrm{freq}$ can be calculated by determining the slope of the line of best fit for a plot of frequency versus $n$ for modes of the same $\ell$. The issue however, is that performing this fit in different ways will alter the values of $\Delta \nu_{\mathrm{freq}}$. For example, the slope of the line will be different depending on which $\ell$ modes are being used. Additionally, there is the question if there should be any weighting on the frequencies. Should modes closer to \numax\ be more heavily weighted? Should the modes be weighted by their uncertainties? Should only modes closest to \numax\ be used? All of these options will result in a different value of $\Delta \nu_{\mathrm{freq}}$.

In the literature there are many different methods used to determine the value of $\Delta \nu$ from individual mode frequencies. For example, \cite{Handberg2017} calculated the average \dnu\ by weighting the frequencies by their observational errors. Other studies have implemented some type of Gaussian weighting around \numax\ \citep{White2011,Rodrigues2017}. \cite{Hekker2013} used both a Gaussian weighted linear fit, an unweighted fit, as well as the median of the pairwise differences of the modes. It is also possible to just use the modes closest to \numax, for example \cite{Corsaro2017AA} determined \dnu\ using a Bayesian linear regression on the asymptotic relation using the central 3 $\ell=0$ mode frequencies.

\begin{table*}
\begin{center}
\caption{A summary of the various different methods used to calculate $\Delta \nu_{\mathrm{freq}}$.}
\label{tabel:Methods_Table}
\begin{tabular}{lll}
\hline
\hline
\textbf{Method} & \textbf{Weighting Used In Slope Determination}                                  & \textbf{Number of Frequencies Used} \\ \hline
\textbf{I}      & None                                                                            & Full Set of Observed Modes Used                                             \\
\textbf{II}     & Error Weighted                                                                  & Full Set of Observed Modes Used                                             \\
\textbf{III}    & Gaussian with a FWHM of 0.66 $\nu_{\mathrm{max}}^{0.88}$ \citep{Mosser2012} & Full Set of Observed Modes Used                                             \\
\textbf{IV}     & None                                                                            & 4 total, 2 on each side of $\nu_{\mathrm{max}}$                         \\
\textbf{V}      & Error Weighted                                                                  & 4 total, 2 on each side of $\nu_{\mathrm{max}}$                         \\
\textbf{VI}     & None                                                                            & 10 total, 5 on each side of $\nu_{\mathrm{max}}$                        \\
\textbf{VII}    & Error Weighted                                                                  & 10 total, 5 on each side of $\nu_{\mathrm{max}}$                        \\ \hline
\end{tabular}
\end{center}
\end{table*}

For each star in our sample, we calculate $\Delta \nu_{\mathrm{freq}}$ in 7 different ways for each $\ell$. So, for stars with $\ell=0,1$ and $2$ values, then $\Delta \nu_{\mathrm{freq}}$ was calculated 21 different times. The different methods of determining \dnu\ are summarized in Table~\ref{tabel:Methods_Table} and explained in more detail as follows:
\begin{enumerate}[I.]
\item{\textbf{No Weighting}: The best-fit slope of the frequency vs. $n$ plot is simply calculated without taking any weighting or uncertainties into account. The full set of observed modes for a given $\ell$ are used.}

\item{\textbf{Error Weighting}: The best-fit slope takes into account the uncertainties in the observed frequency values. The full set of observed modes for a given $\ell$ are used.}

\item{\textbf{Gaussian Weighting}: High signal-to-noise data shows that the power envelope of the excited modes is a Gaussian with a full-width-half-max of $0.66 \nu_{\mathrm{max}}^{0.88}$ \citep{Mosser2012}. Therefore, it may be reasonable to weight those modes closer to \numax\ more heavily in the best-fit slope. Here each frequency is given a weight, where the weighting function is a Gaussian centered on \numax\ with a FWHM given by $0.66 \nu_{\mathrm{max}}^{0.88}$. The weight, $W$, for each point is then given by
\begin{equation}
W = e^{-(\nu - \nu_\mathrm{max})^2 / (2 \sigma^2)}
\end{equation} 
where $\sigma=\mathrm{FWHM} / (2 \sqrt{2 \ln(2)})$. This is similar to what was done in \cite{Rodrigues2017}, with the difference being the value of $\sigma$ used. \cite{Rodrigues2017} use $\sigma = 0.66 \nu_{\mathrm{max}}^{0.88}$, while we use that as our FWHM, thus making our values of $\sigma$ differ by a factor of $2 \sqrt{2 \ln(2)}$. With the weighting for each frequency determined, the slope of the line of best fit is then calculated by minimizing
\begin{equation}
\sum_{i=1}^{k} W_{i} [\nu_{i} - (\mathrm{slope} \times n_{i} + \mathrm{intercept})]^2,
\end{equation}
where $k$ is the number of modes. The full set of observed modes for a given $\ell$ are used.}

\item{\textbf{No Weighting, 4 Points}: Only 4 frequencies are used in the fit, 2 on each side of \numax. No errors or weighting equation is used.}

\item{\textbf{Error Weighting, 4 Points}: Only 4 frequencies are used in the fit, 2 on each side of \numax. The uncertainties in the observed frequencies are included in the best-fit slope calculation.}

\item{\textbf{No Weighting, 10 Points}: Only 10 frequencies are used in the fit, 5 on each side of \numax. No errors or weighting equation is used.}

\item{\textbf{Error Weighting, 10 Points}: Only 10 frequencies are used in the fit, 5 on each side of \numax. The uncertainties in the observed frequencies are included in the best-fit slope calculation.}

\end{enumerate}
For each star, and for each $\ell$, these 7 methods were used to compute $\Delta \nu_{\mathrm{freq}}$. For the remainder of the paper the methods will be referred to by their corresponding Roman numeral. Note that when determining the individual mode frequencies in actual stellar models then methods II, V, and VII cannot be used since there is no associated observational uncertainty on the frequencies.

\section{Results}
\label{Sec:Results}
\subsection{Comparing $\Delta \nu_{\mathrm{CV}}$ and $\Delta \nu_{\mathrm{freq}}$}
\label{Sec:Obs_results}
\begin{figure*}
\epsscale{1.0}
\plotone{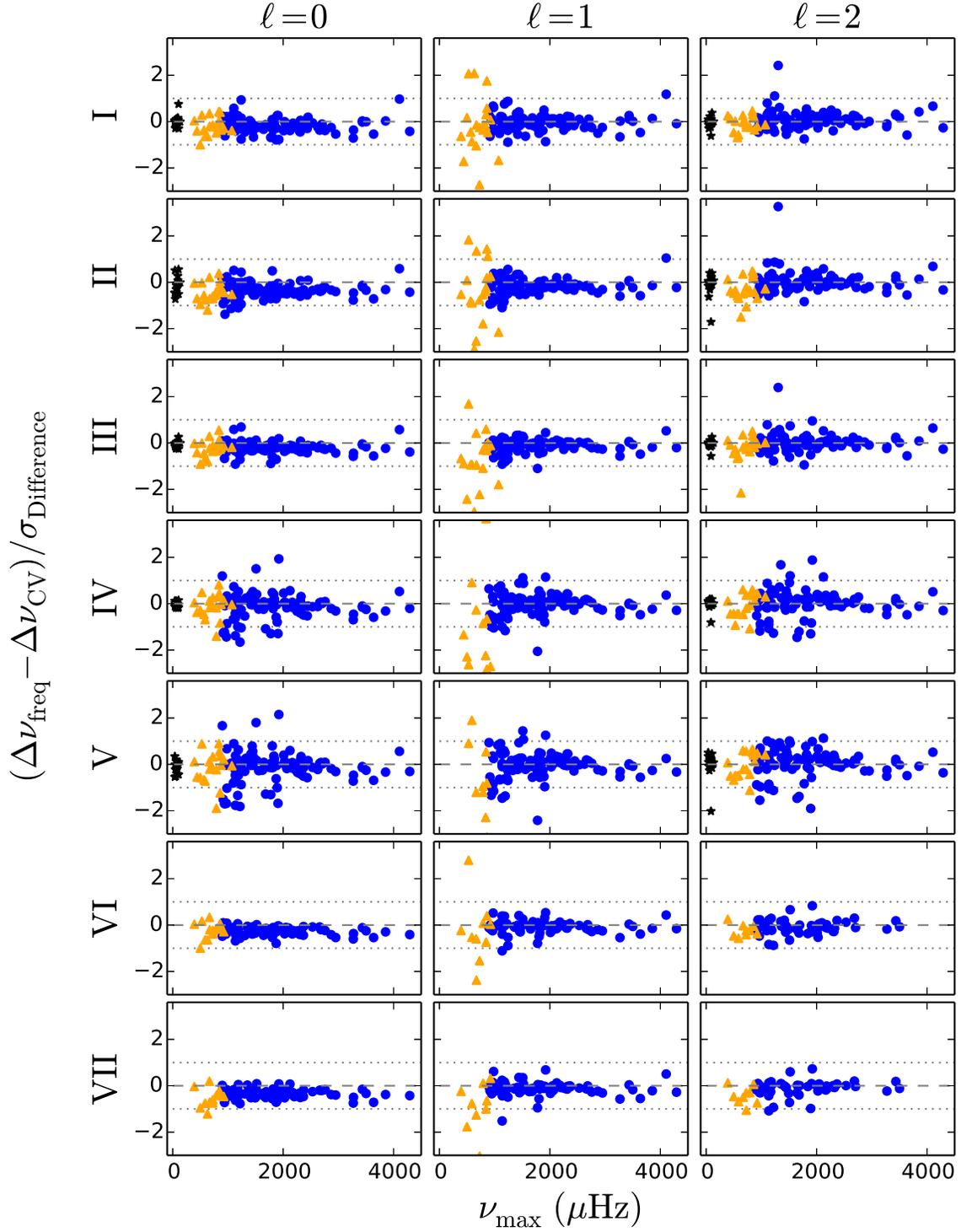}
\caption{The difference,  $(\Delta \nu_{\mathrm{freq}} - \Delta \nu_{\mathrm{CV}})  / \sigma_{\mathrm{Difference}}$, as a function of \numax\ for each of the different methods of determining $\Delta \nu_{\mathrm{freq}}$ and for each $\ell$. Note that some of the subgiant stars are not in the ordinate range of this plot. Colors are the same as Fig.~\ref{fig:ACF_Bell_dnu_compare_plot}. For reference, the dashed line is at 0 and the dotted lines are at $\pm$1.}
\label{fig:frac_diff_dnu}
\end{figure*}

The large frequency spacing calculated using the CV method, $\Delta \nu_{\mathrm{CV}}$, can be compared to the different values of $\Delta \nu_{\mathrm{freq}}$ calculated in Sec.~\ref{Sec:dnu_freq_calc}. Figure~\ref{fig:frac_diff_dnu} shows the difference between the values of $\Delta \nu_{\mathrm{freq}}$ and $\Delta \nu_{\mathrm{CV}}$ divided by the uncertainty in the difference, as a function of \numax, for each different method of calculating $\Delta \nu_{\mathrm{freq}}$ for the $\ell=0,1,$ and $2$ modes. As can be seen in Figure~\ref{fig:frac_diff_dnu}, the difference between the values of \dnu\ are mostly all within 1$\sigma$. The exception to this is \dnu\ calculated with the $\ell=1$ modes in the subgiant stars, which have a large scatter. This is due to the fact that the subgiant stars have mixed-modes. For the sake of making the ordinate scale in Fig.~\ref{fig:frac_diff_dnu} small enough to easily view the data, some of the values of \dnu\ calculated using the $\ell=1$ frequencies for the subgiant stars fall outside the range of the figure and are not visible. Additionally, one should note that for the RGB stars there were not enough modes on either side of \numax\ for the value of $\Delta \nu_{\mathrm{freq}}$ to be calculated using methods VI and VII. 

It can also be seen in Figure~\ref{fig:frac_diff_dnu} that there is a larger scatter and disagreement for the method of determining $\Delta \nu_{\mathrm{freq}}$ using only 2 frequencies on either side of \numax\ (methods IV and V). This suggests that using only 4 frequencies in the \dnu\ determination is not ideal when attempting to match the value of \dnu\ one would calculate from the observed power spectrum. This larger scatter make sense due to the fact that the value of $\Delta \nu_{\mathrm{CV}}$ uses nearly the entire region of oscillations from the power spectrum and therefore takes many more than 4 frequencies into account. So, one might expect that the value of $\Delta \nu_{\mathrm{CV}}$ and $\Delta \nu_{\mathrm{freq}}$ would match more poorly when $\Delta \nu_{\mathrm{freq}}$ only uses a few modes.

Additionally, for some of the methods it does appear that the value of $\Delta \nu_{\mathrm{freq}}$ tends to be smaller than the value of $\Delta \nu_{\mathrm{CV}}$. This can be seen in Fig.~\ref{fig:frac_diff_dnu} for the $\ell=0$ modes in methods II, III, VI, and VII as well as the $\ell=1$ and $\ell=2$ modes for method VII. Despite the differences among the methods, with exception to some of the subgiant stars, the value of $\Delta \nu_{\mathrm{CV}}$ and $\Delta \nu_{\mathrm{freq}}$ agree to within a few percent. 

\subsection{Simulated TESS Data}
\begin{figure}[h]
\epsscale{1.0}
\plotone{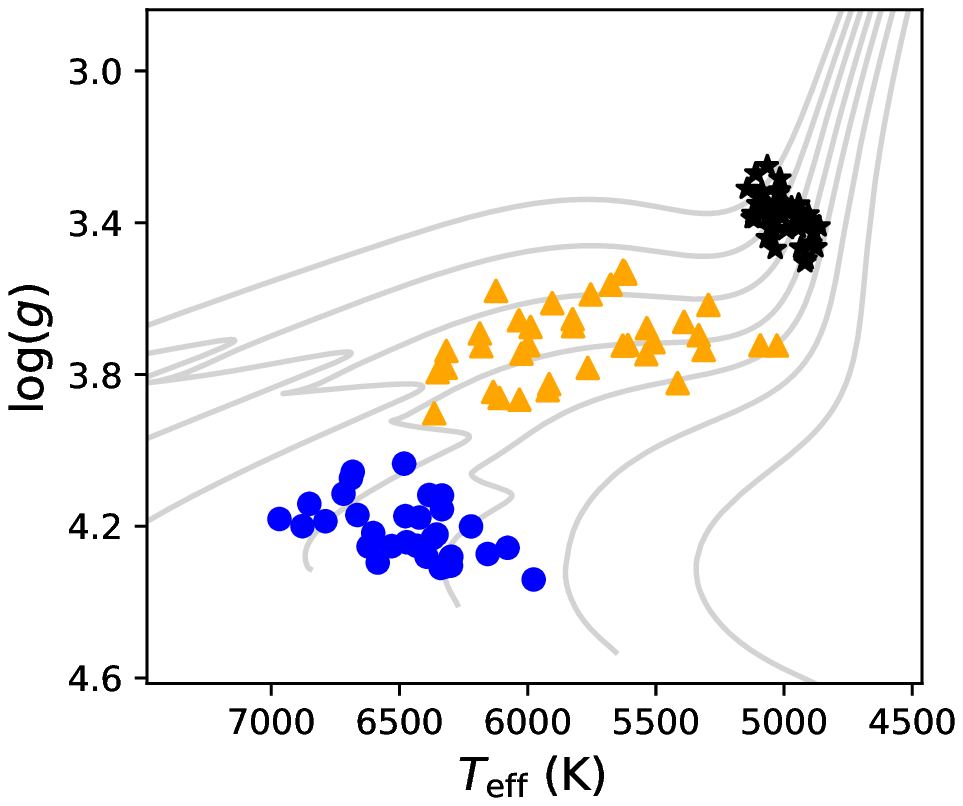}
\caption{Kiel diagram of the stars selected from \cite{Ball2018}. The black points are the red giant stars, the orange triangles are the subgiant stars, and the blue circles are the main sequence stars. The background gray lines show tracks of mass 0.8, 1.0, 1.2, 1.4, 1.6, 1.8, and 2.0 $M_\sun$, created using YREC \citep{Demarque2008}.}
\label{fig:TESS_selections}
\end{figure}

So far our comparisons have used observed data with the determined mode frequencies acting as a proxy for the frequencies one would have from a stellar model. Since the properties of the observations, for example the S/N or time-series length, will determine which modes were observed, we need to make sure that our results also hold when using stellar models. Additionally, since the goal of this project is to determine the best method of calculating \dnu\ from stellar models, it is critical that we repeat the experiment using frequencies from actual stellar models. To accomplish this, we make use of the simulated TESS data from \cite{Ball2018}. \cite{Ball2018} created a mock catalog of lightcurves to simulate data from NASA's \textit{Transiting Exoplanet Survey Satellite}, TESS \citep{Ricker2015}. From this mock catalog we selected 34 main sequence stars, 37 subgiants, and 47 red giants to analyze. The selected stars can be seen in the Kiel diagram in Figure~\ref{fig:TESS_selections}. Note that the \cite{Ball2018} simulated TESS data was restricted to stars which would be observed with the short cadence (2 minutes) mode of TESS and as a result the red giant stars available in this catalog are those which are not evolved too far along the red giant branch. All of the selected stars and lightcurves used are from ``Sector 1" in \cite{Ball2018}. Additionally, it should be noted that the lightcurves used did not have white noise added to them. While \cite{Ball2018} does provide the expected value of the white noise for each lightcurve, the scope of this project is not concerned with the actual observing capabilities of TESS, rather the comparison between the value of $\Delta \nu$ from the power spectrum and $\Delta \nu$ from frequencies. Hence, the clean lightcurves from \cite{Ball2018} were used without added noise.

From the simulated TESS data, the ``observed'' values were calculated from the lightcurves. The values of \numax\ and $\Delta \nu_{\mathrm{CV}}$ were calculated using the coefficient of variation method as discussed in Sec.~\ref{Sec:CV_method}. For the simulated stars \cite{Ball2018} also models each star and provides individual mode frequencies. It should be noted that the frequencies from the models were used in the creation of the simulated spectra. So, unlike typical model frequencies which may disagree with the true frequency values, these modeled frequencies are actually those found in the simulated spectra. Using these model frequencies, the value of $\Delta \nu_{\mathrm{freq}}$ was calculated as discussed in Sec.~\ref{Sec:dnu_freq_calc}. The frequencies provided from the \cite{Ball2018} models are for values between 0.15 and 0.95$\nu_\mathrm{ac}$. Thus, for the model stars, when calculating $\Delta \nu_\mathrm{freq}$ using methods I and III, which utilize the full set of modes, this corresponds to all modes between 0.15 and 0.95$\nu_\mathrm{ac}$.

\begin{figure*}
\epsscale{1.0}
\plotone{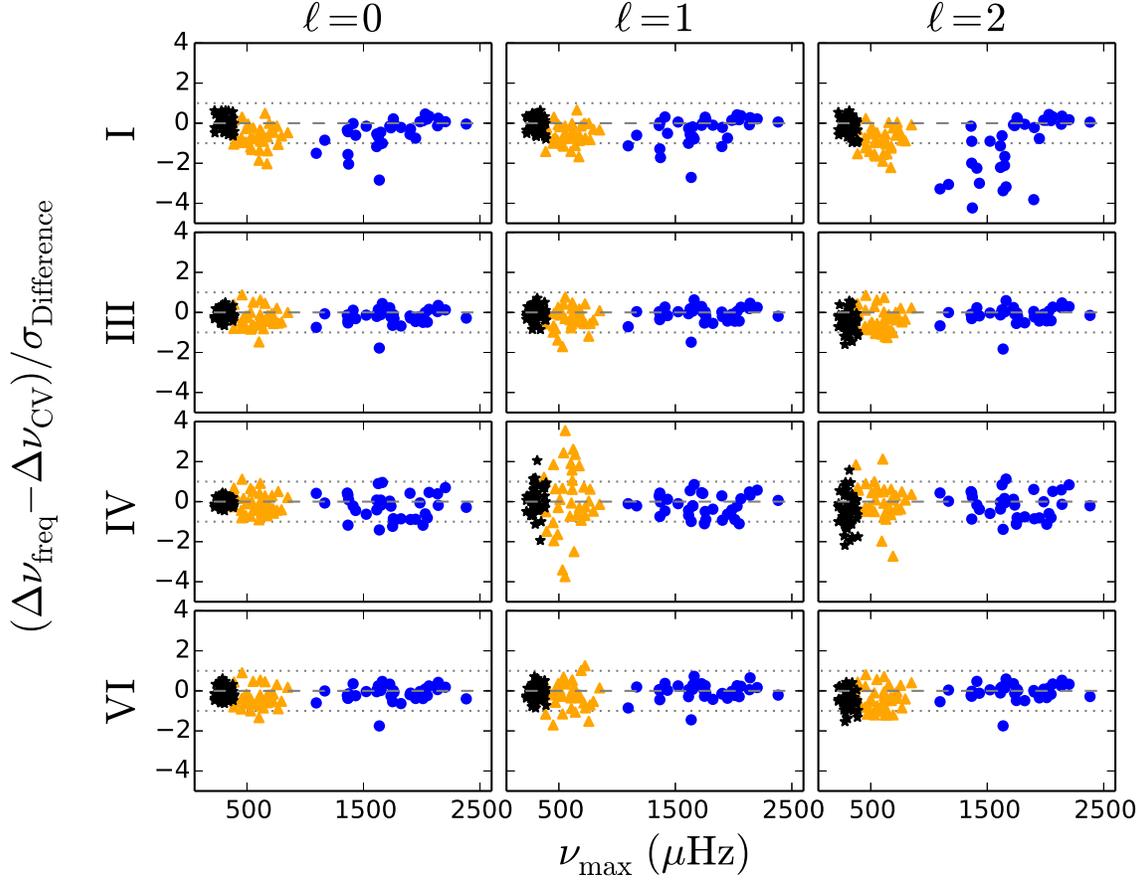}
\caption{The difference, $(\Delta \nu_{\mathrm{freq}} - \Delta \nu_{\mathrm{CV}})  / \sigma_{\mathrm{Difference}}$, as a function of \numax\ for each of the different methods of determining $\Delta \nu_{\mathrm{freq}}$ and for each $\ell$ for stars from the TESS simulated data. Colors are the same as Fig.~\ref{fig:ACF_Bell_dnu_compare_plot}. For reference, the dashed line is at 0 and the dotted lines are at $\pm$1.}
\label{fig:frac_diff_dnu_TESS}
\end{figure*}

As in Sec.~\ref{Sec:Obs_results}, we then compare the values of $\Delta \nu_{\mathrm{CV}}$ and $\Delta \nu_{\mathrm{freq}}$. Figure~\ref{fig:frac_diff_dnu} can be remade, but for the sample of stars from the simulated TESS data. This can be seen in Figure~\ref{fig:frac_diff_dnu_TESS}. Note that methods II, V, and VII are not used in this case. This is due to the fact that these methods use the observational uncertainties when determining the value of $\Delta \nu_{\mathrm{freq}}$ and the model values do not have observational uncertainties. As seen in Fig.~\ref{fig:frac_diff_dnu_TESS}, again the values of $\Delta \nu_{\mathrm{CV}}$ and $\Delta \nu_{\mathrm{freq}}$ agree well and for the most part are within 1 or 2$\sigma$. Additionally, it can be seen that for method I the value of $\Delta \nu_{\mathrm{freq}}$ tends to be smaller than the value of $\Delta \nu_{\mathrm{CV}}$. 

\section{Discussion}
\label{Sec:Discussion}
We can put the information in Figures \ref{fig:frac_diff_dnu} and \ref{fig:frac_diff_dnu_TESS} on a more quantitative footing to determine which method of calculating $\Delta \nu_{\mathrm{freq}}$ is most in agreement with the value of $\Delta \nu_{\mathrm{CV}}$. The root mean square (rms) value of the percent difference, $ 100 \times (\Delta \nu_{\mathrm{freq}} - \Delta \nu_{\mathrm{CV}}) / \Delta \nu_{\mathrm{CV}}$, can be compared for each method of calculating $\Delta \nu_{\mathrm{freq}}$. Table~\ref{table:rms_all_1} shows the rms value for each method of calculating $\Delta \nu_{\mathrm{freq}}$ for each set of stars for the observed sample. As can be seen in Table~\ref{table:rms_all_1}, the best method to calculate $\Delta \nu_{\mathrm{freq}}$ depends on the star's evolutionary stage and the $\ell$ of interest.

For the main sequence stars using the $\ell=0$ modes, methods I, III, and VI perform equally well and produce values of $\Delta \nu_{\mathrm{freq}}$ closer to the values of $\Delta \nu_{\mathrm{CV}}$ than the other methods. Using the $\ell=1$ modes instead, methods I, III, and VI again outperform the other methods, with method VII being equally good as well. When using only the $\ell=2$ modes methods VI and VII perform equally well and better than the other methods. However, we must be careful with comparing all the methods at once, since not every star in the sample can be included in every method of calculating $\Delta \nu_{\mathrm{freq}}$. For example, not all the stars had enough observed modes to use methods VI and VII. Since all the stars are included in methods I, II, and III, then it is safest to compare these three against each other. Doing this we see that for the main sequence stars using the $\ell=0$ or $\ell=1$ modes are much better than using the $\ell=2$ modes. Also, we see that using methods I and III are nearly equivalent and better than method II. So, this means that for the main sequence stars using either no weighting or the \cite{Mosser2012} Gaussian weighting is better than weighting by the observational uncertainties.

For the subgiant stars, using the $\ell=0$ modes, methods I, III, and VI perform nearly equivalently and are better than the other methods. The $\ell=1$ modes do not provide good values, as expected due to the presence of mixed-modes. For the $\ell=2$ modes, all the methods perform about the same except for methods I and VII being worse.

For the red giant stars, using the $\ell=0$ modes, method III is the best, however all methods except II do equally well. Note that methods VI and VII are not included for the RGB stars because there were not enough modes to have 5 frequencies on each side of \numax. Using the $\ell=2$ modes, methods I and III perform equally well and better than the other methods. For every method, using the $\ell=0$ modes provided the best results for the RGB stars. Similar to the main sequence stars, we see that using either no weighting or the \cite{Mosser2012} Gaussian weighting is better than using the observational uncertainties.

\begin{table*}
\caption{The rms value for the percent difference, $ 100 \times (\Delta \nu_{\mathrm{freq}} - \Delta \nu_{\mathrm{CV}})  / \Delta \nu_{\mathrm{CV}}$, for each method and each $\ell$ for our set of \textit{Kepler} stars.}
\label{table:rms_all_1}
\begin{tabular}{llll|lll|lll|lll|lll}
\hline
\hline
        & \multicolumn{3}{c|}{$\ell=0$}                 & \multicolumn{3}{c|}{$\ell=1$}         & \multicolumn{3}{c|}{$\ell=2$}                 & \multicolumn{3}{c|}{Average  of $\ell=0,1$}        & \multicolumn{3}{c}{Average  of $\ell=0,2$}         \\ \hline
        & MS          & Subgiant   & RGB         & MS          & Subgiant   & RGB & MS          & Subgiant   & RGB         & MS          & Subgiant   & RGB         & MS          & Subgiant   & RGB         \\ \hline
I    & 0.37 & 0.81 & 0.81 & 0.36 & 1.84 & -- & 0.63 & 1.63 & 0.91 & 0.33 & 1.16  & 0.81 & 0.38 & 1.10 & 0.75 \\
II   & 0.53 & 1.06 & 0.94 & 0.40 & 2.73 & -- & 0.63 & 1.04 & 1.25 & 0.44 & 1.68  & 0.94 & 0.44 & 1.02 & 0.93 \\
III    & 0.36 & 0.88 & 0.72 & 0.35 & 2.34 & -- & 0.63 & 0.97  & 0.92  & 0.29 & 1.50 & 0.72 & 0.39 & 0.90 & 0.72 \\
IV & 0.65  & 1.09 & 0.82 & 0.54 & 11.74 & -- & 0.64 & 1.01 & 1.43  & 0.55 & 5.90 & 0.82 & 0.59 & 0.95  & 0.72 \\
V & 0.65 & 1.16  & 0.76 & 0.54 & 12.33 & -- & 0.64 & 0.91 & 1.41 & 0.56 & 6.20 & 0.76 & 0.60 & 0.96 & 0.57  \\
VI & 0.34 & 0.86 & --    & 0.29 & 3.18 & -- & 0.28 & 0.91 & --         & 0.27 & 1.73 & --         & 0.32 & 0.81 & --         \\
VII & 0.41 & 1.17 & --   & 0.30 & 3.78   & -- & 0.29 & 1.24 & --         & 0.33  & 2.18 & --   & 0.36 & 1.13 & --         \\ \hline
\end{tabular}
\end{table*}

\begin{table*}
\caption{The rms value for the percent difference, $100 \times (\Delta \nu_{\mathrm{freq}} - \Delta \nu_{\mathrm{CV}})  / \Delta \nu_{\mathrm{CV}}$, for each method for the simulated TESS stars.}
\label{table:rms_TESS}
\begin{tabular}{llll|lll|lll|lll|lll}
\hline
\hline
        & \multicolumn{3}{c|}{$\ell=0$}                 & \multicolumn{3}{c|}{$\ell=1$}         & \multicolumn{3}{c|}{$\ell=2$}                 & \multicolumn{3}{c|}{Average  of $\ell=0,1$}        & \multicolumn{3}{c}{Average  of $\ell=0,2$}         \\ \hline
        & MS          & Subgiant   & RGB         & MS          & Subgiant   & RGB & MS          & Subgiant   & RGB         & MS          & Subgiant   & RGB         & MS          & Subgiant   & RGB         \\ \hline
I    & 1.04 & 1.22 & 0.59 & 0.96 & 1.12 & 0.59 & 	1.91 & 1.49 & 0.71 & 1.00 & 	1.15 	& 0.57 & 1.43 & 1.34	 & 0.62  \\
III    &  0.61 & 0.97 & 0.60 & 0.52 & 1.08 & 0.73 & 0.60 & 1.10 & 1.38 & 	0.56 & 0.89 & 0.63 & 0.60 & 1.00 & 0.93   \\
IV & 0.83 & 1.03 & 0.64 & 0.67 & 3.51 & 1.84 & 0.81 & 1.76 & 2.53 & 0.73 & 1.72 & 1.07 & 0.82 & 1.18 & 1.44  \\
VI &  0.58 & 0.99 & 	0.62	 & 0.51 & 1.22 & 0.66 & 0.57	 & 1.18 & 1.16 & 0.54 & 0.96 & 0.61 & 0.57 & 1.06	 & 0.82    \\ \hline
\end{tabular}
\end{table*}

We can perform the same investigation for the simulated TESS data, again looking at the rms value of the percent difference, $100 \times (\Delta \nu_{\mathrm{freq}} - \Delta \nu_{\mathrm{CV}}) / \Delta \nu_{\mathrm{CV}}$, for each method. Table~\ref{table:rms_TESS} shows the rms value for each method of calculating $\Delta \nu_{\mathrm{freq}}$ for each set of stars and each $\ell$. There are a few important differences to note with using the simulated TESS data compared to the observational data. First of all, for the simulated TESS data, we only selected the $\ell=1$ modes which were not mixed-modes. So, the rms value of the subgiant $\ell=1$ modes are unrealistically good. Additionally, since the mode frequencies are from models, we have many more modes than we would actually have from observations for these stars. This also means that all of the stars in our sample of simulated TESS data had enough modes that all stars could be put through each method of calculating $\Delta \nu_{\mathrm{freq}}$ for every $\ell$. However, since the modes are from stellar models, methods II, V, and VII could not be calculated since there was no observational uncertainty on the modes. 

As can be seen in Table~\ref{table:rms_TESS}, for the main sequence stars, when using the $\ell=0$ modes, methods III and VI perform nearly equally well and better than the other methods. Methods III and VI again outperform the others when using the $\ell=1$ modes or the $\ell=2$ modes as well. When comparing using the $\ell=0$ and $\ell=2$ modes for the main sequence stars we see that for methods III, IV, and VI the resulting rms values are very similar. For method I using the $\ell=0$ modes is better than using the $\ell=2$ modes.

For the subgiant stars methods III and VI are slightly better for the $\ell=0$ modes. For the $\ell=1$ and $
\ell=2$ modes method III performs the best. For every method, using the $\ell=0$ modes for the subgiants outperforms using the $\ell=2$ modes.

For the red giant branch stars, regardless of which method is being used, using the $\ell=0$ modes give a lower rms value than using the $\ell=1$ or $\ell=2$ modes. Additionally, regardless of which $\ell$ modes are being used, method I is the best in all cases. However, since for these simulated stars we have a lot more modes than would get in observations, then perhaps using all available frequencies as in method I is unrealistic. If instead for the RGB stars we are restricted to only using 5 points on either side of \numax, as in method VI, then the resulting rms values are either better or nearly the same as using the Gaussian weighting of method III. 

Finally, for our observed RGB stars we can compare our $\nu_{\mathrm{max,CV}}$ and $\Delta \nu_{\mathrm{CV}}$ values to those from \cite{Corsaro2017AA}. \cite{Corsaro2017AA} determined \dnu\ using a Bayesian linear regression on the asymptotic relation using the central 3 $\ell=0$ mode frequencies. We see that the values of \dnu\ agree to within 1.5\% and values of \numax\ agree to within 10\%. Comparing the values of $\Delta \nu_{\mathrm{freq}}$ to the \dnu\ values from \cite{Corsaro2017AA} we see that for the $\ell=0$ modes, method V is in the most agreement. This is as expected since method V uses 2 points on each side of \numax\ with error weights and is the most similar method compared to the technique of using the 3 central frequencies performed in \cite{Corsaro2017AA}.

\section{Summary and Conclusion}
\label{Sec:Conclusion}
When comparing observed values of \dnu\ to values of \dnu\ calculated from stellar models, we must be aware that the manner in which these two values of \dnu\ are being determined are different. The observed value of $\Delta \nu$ from photometric time series data is typically determined using autocorrelation techniques or PS$\otimes$PS methods, while this is not the case when calculating \dnu\ from stellar models. In stellar models the actual individual mode frequencies are calculated and then the value of $\Delta \nu_{\mathrm{freq}}$ can be determined by fitting a line to the frequency versus $n$ data. There are many different methods by which to perform this linear fit, for example to weight the frequencies closer to \numax\ more heavily, to only use a few points around \numax, and deciding which $\ell$ modes to include. It is critical that when we determine the value of \dnu\ for stellar models that we are doing so in a way that will provide consistent results with the values of \dnu\ calculated through observations. Otherwise the comparison between the two values of \dnu\ looses accuracy.

In this work we took high signal-to-noise \textit{Kepler} observations and determined the seismic parameters using standard methods. Each of these stars also had individual mode frequencies determined in the literature. Using these individual mode frequencies as a proxy for the frequencies one would have from modeling a star, we determined $\Delta \nu_{\mathrm{freq}}$ in several different ways to compare to the value of $\Delta \nu_{\mathrm{CV}}$. We also made use of simulated TESS lightcurves and stellar models to compare the methods of calculating \dnu\ values. From the results of comparing $\Delta \nu_{\mathrm{CV}}$ and $\Delta \nu_{\mathrm{freq}}$ both from the observed stars and the simulated TESS data, we show that using the $\ell=0$ modes with either no weighting or a Gaussian weighting as in \cite{Mosser2012} provides the best agreement. Additionally, we see that using the coefficient of variation method as in \cite{Bell2018} provides a quick and accurate way to identify the frequency range of excited modes.

\acknowledgments This work was partially supported by NSF grant AST-1514676 and NASA grant NNX16AI09G to SB. EC is funded by the European Union's Horizon 2020 research and innovation program under the Marie Sklodowska-Curie grant agreement No. 664931. WHB and WJC acknowledge support from the UK Science and Technology Facilities Council (STFC). WHB acknowledges support from the UK Space Agency (UKSA). Funding for the Stellar Astrophysics Centre is provided by The Danish National Research Foundation (grant DNRF106). The authors thank the anonymous referee for the helpful comments and suggestions.

\software{\textsc{Diamonds} \citep{Corsaro2014} and YREC \citep{Demarque2008}}

\bibliography{dnu_paper}

\appendix
\section{Testing the Various CV Methods}
\label{Sec:Appendix}
As mentioned in Sec~\ref{Sec:CV_method}, many variations of the CV method were tested. These different implementations of the CV method will be described here, where we refer to the method described in Sec~\ref{Sec:CV_method} as the ``base'' method. In the ``base'' method, each step we shift the window towards higher frequencies by $1/6$ the previous window size. We also tested shifting the window by $1/4$, $1/2$, $1/10$, and $1/20$ of the previous window size. This effectively changes the density of the blue points in Fig.~\ref{fig:CV_plot}. Also, we tested using different window sizes when calculating the CV value along the spectrum. In the base implementation the window size is determined by estimating the value of $\Delta \nu$ at the central frequency. We also tested window sizes of 2$\Delta \nu$, 4.2$\Delta \nu$, and $W_\mathrm{Mosser}$. Additionally, when smoothing the CV values the size of the smoothing window was tested. In the base method we smoothed with a window of width $W_{\mathrm{Mosser}}$ and here we test using a window of width 2$W_{\mathrm{Mosser}}$, $W_{\mathrm{Mosser}}/2$, $2 \Delta \nu$, $4.2 \Delta \nu$, $6 \Delta \nu$, and $8 \Delta \nu$. The final parameter that was tested was the window sizes used when determining the weighted centroid of the peak. In the base implementation we use a window of $W_{\mathrm{Mosser}}$ and here we also test a window of $2 W_{\mathrm{Mosser}}$.

To test which method works best a set of 63 randomly selected red giant stars from the Second APOKASC Catalog \citep{Pinsonneault2018} were used. Using the power spectrum from KASOC, the \numax\ values of the 63 stars were calculated using the base CV method as described in Sec.~\ref{Sec:CV_method}. Then the various different implementations of the CV method, described in the previous paragraph, were used to again calculate $\nu_\mathrm{max}$. The resulting \numax\ values were compared to the values of \numax\ from the Second APOKASC paper \citep{Pinsonneault2018}, as well as the different pipelines in the Second APOKASC Catalog: A2Z \citep{Mathur2010,Garcia2014}, CAN \citep{Kallinger2010}, COR \citep{Mosser2009}, OCT \citep{Hekker2010}, and SYD \citep{Huber2009}. The rms value for the percent difference, $100 \times (\nu_\mathrm{max,pipeline} - \nu_\mathrm{max,CV}) / \nu_\mathrm{max,CV}$, was calculated for each CV implementation and can be seen in Table~\ref{table:CV_RGB_rms_tests}. While the ``best'' method depends on which pipeline the \numax\ value is being compared to, the base CV method was in good agreement across all sets. Additionally, if the rms values for the pipelines are averaged together, as seen in the last column of Table~\ref{table:CV_RGB_rms_tests}, then the base CV method performs the best. Therefore, this base method was the one selected as the best implementation of the CV method and is the one used throughout the paper. 

\begin{table*}[h]
\begin{center}
\caption{The rms value for the percent difference, $100 \times (\nu_\mathrm{max,pipeline} - \nu_\mathrm{max,CV}) / \nu_\mathrm{max,CV}$ for each CV implementation for our sample of RGB stars from the Second APOKASC Catalog \citep{Pinsonneault2018}. The columns represent the various $\nu_\mathrm{max}$ pipeline values from the catalog. The last column is an average of the rms value for all the pipelines.}
\label{table:CV_RGB_rms_tests}
\begin{tabular}{l|rrrrrr|r}
\hline
\hline
                   & \multicolumn{6}{c|}{\textbf{Pipeline}}                                                                                                                                                                               & \multicolumn{1}{l}{}                 \\ \hline
\textbf{CV Method}    & \multicolumn{1}{l}{\textbf{APOKASC}} & \multicolumn{1}{l}{\textbf{A2Z}} & \multicolumn{1}{l}{\textbf{CAN}} & \multicolumn{1}{l}{\textbf{COR}} & \multicolumn{1}{l}{\textbf{OCT}} & \multicolumn{1}{l|}{\textbf{SYD}} & \multicolumn{1}{l}{\textbf{Average}} \\ \hline
Base Method               & 2.83                                 & 3.20                             & 3.10                             & 3.20                             & 2.97                             & 3.02                              & 3.05                                 \\
Window Shift: $1/4$ Previous               & 3.13                                 & 3.38                             & 3.44                             & 3.53                             & 3.33                             & 3.21                              & 3.34                                 \\
Window Shift: $1/2$ Previous               & 3.73                                 & 3.97                             & 3.96                             & 4.17                             & 3.72                             & 3.85                              & 3.90                                 \\
Window Shift: $1/10$ Previous              & 3.16                                 & 3.67                             & 3.36                             & 3.57                             & 2.94                             & 3.44                              & 3.35                                 \\
Window Shift: $1/20$ Previous              & 2.91                                 & 3.40                             & 3.05                             & 3.34                             & 2.80                             & 3.23                              & 3.12                                 \\
Window Size: 2$\Delta \nu$            & 4.56                                 & 4.71                             & 4.26                             & 5.00                             & 4.76                             & 4.85                              & 4.69                                 \\
Window Size: 4.2$\Delta \nu$          & 12.98                                & 12.69                            & 12.52                            & 13.28                            & 13.56                            & 13.08                             & 13.02                                \\
Window Size: $W_\mathrm{Mosser}$          & 13.15                                & 12.85                            & 12.60                            & 13.41                            & 13.81                            & 13.29                             & 13.19                                \\
Smoothing Window: 2$\Delta \nu$      & 3.07                                 & 3.19                             & 3.20                             & 3.50                             & 3.44                             & 3.24                              & 3.27                                 \\
Smoothing Window: 4.2$\Delta \nu$        & 2.89                                 & 3.45                             & 2.97                             & 3.26                             & 2.83                             & 3.24                              & 3.11                                 \\
Smoothing Window: 6$\Delta \nu$      & 4.92                                 & 5.02                             & 4.69                             & 5.42                             & 4.97                             & 5.23                              & 5.04                                 \\
Smoothing Window: 8$\Delta \nu$      & 10.91                                & 10.90                            & 10.36                            & 11.18                            & 11.20                            & 11.22                             & 10.96                                \\
Smoothing Window: 2$W_\mathrm{Mosser}$   & 10.45                                & 10.62                            & 10.16                            & 10.71                            & 10.32                            & 10.77                             & 10.51                                \\
Smoothing Window: $W_\mathrm{Mosser} / 2$ & 3.00                                 & 3.06                             & 3.29                             & 3.42                             & 3.40                             & 3.07                              & 3.21                                 \\
Weighted Centroid Peak Width: 2$W_\mathrm{Mosser}$         & 3.33                                 & 3.92                             & 4.05                             & 3.52                             & 2.81                             & 3.41                              & 3.51                                 \\ \hline
\end{tabular}
\end{center}
\end{table*}

\end{document}